\documentclass[hyper]{JHEP3}
\usepackage[dvips]{graphicx}
\usepackage{latexsym,amsmath,amssymb}
\usepackage{epsfig,subfigure}

\title{Particle Collisions on Stringy Black Hole Background}

\author{Shao-Wen Wei,
        Yu-Xiao Liu\footnote{Corresponding author.},
        Hai-Tao Li,
        Feng-Wei Chen\\
   Institute of Theoretical Physics, Lanzhou University,
           Lanzhou 730000, P. R. China \\
  E-mail: \email{weishaow06@lzu.cn}
          \email{liuyx@lzu.edu.cn}
          \email{liht07@lzu.cn}
          \email{chenfw06@lzu.cn}}

\date{\today}

\abstract{%
The collision of two particles in the background of a Sen
black hole is studied. With the equations of motion of the particles,
the center-of-mass energy is investigated when the collision takes
place at the horizon of a Sen black hole. For an extremal Sen black hole, we find that the center-of-mass energy will be arbitrarily high with two conditions: (1) spin $a\neq 0$ and (2) one of the colliding particles has the critical angular momentum $l_{\text{c}}=2$. For a nonextremal Sen black hole, we show that, in order to obtain an unlimited center-of-mass energy, one of the colliding particles should have the critical angular momentum $l'_{\text{c}}=2 r_{+}/a$ ($r_{+}$ is the radius of the outer horizon for a nonextremal black hole). However, a particle with the angular momentum $l=l'_{\text{c}}$ could not approach the black hole from outside of the horizon through free fall, which implies that the collision with arbitrarily high center-of-mass energy could not take place. Thus, there is an upper bound of the center-of-mass energy for the nonextremal black hole. We also obtain the maximal center-of-mass energy for a near-extremal black hole and the result implies that the Planck-scale energy is hard to be approached. Furthermore, we also consider the back-reaction effects. The result shows that, neglecting the gravitational radiation, it has a weak effect on the center-of-mass energy. However, we argue that the maximum allowed center-of-mass energy will be greatly reduced to below the Planck-scale when the gravitational radiation is included.}


\begin{document}

\section{Introduction}
\label{secIntroduction}

With the help of the largest terrestrial accelerator,
the Large Hadron Collider (LHC), one could detect the physics at
collision energy $10$ TeV. However, comparing with the
Planck-scale energy $10^{16}$ TeV, it is too low to probe the
Planck-scale physics. So, other new physical mechanisms should be
proposed for the purpose that probing the Planck-scale physics, and
which may also contribute to the discovery of extra dimensions of
spacetime and the Grand Unification theory.

Fortunately, Ba\~nados, Silk and West (BSW) \cite{Banados} recently
suggested that the spinning Kerr black holes could play the role of
particle accelerators. Especially, the center-of-mass (CM) energy
for a pair neutral particle (for example, the dark matter particles) colliding at the horizon of an extremal black hole can be unlimited. The property of the collision will provide a unique probe of the Planck-scale physics. However, the
fine-tunings arise, namely, the black hole must be an extremal one and one of the colliding particles should have orbital
angular momentum per unit rest mass $l=2$. Whereafter, in refs. \cite{Berti} and
\cite{Jacobson}, the authors further elucidated the mechanism for
the result of BSW and argued that there must exist a practical
limitation on the achievable CM energy for the reason that there
always exists a small deviation of the spin of the astrophysical
black hole from its maximal value \cite{Thorne}. On the other hand, Lake studied the
CM energy of the collision taking place at the inner horizon of a
nonextremal Kerr black hole and the CM energy was found to be
limited \cite{Lake}. Grib and Pavlov \cite{Grib1,Grib2,Grib3}
investigated the particle collisions and the extraction of energy in
the background of a Kerr black hole. The universal property of
acceleration of particles for a rotating black hole is discussed in
\cite{Zaslavskii}. In our previous work \cite{Wei}, we studied the
property of the CM energy for two colliding particles in the
background of a Kerr-Newman (KN) black hole. We showed that the CM
energy can be arbitrarily high with the conditions: (1) the
collision takes place at the horizon of an extremal black hole; (2)
one of the colliding particles has critical angular momentum; (3) the
spin $a$ of the extremal black hole satisfies
$\frac{1}{\sqrt{3}}\leq \frac{a}{M}\leq 1$, where $M$ is the mass of the black hole.
In this paper, we shall study the property of the collision of two
uncharged particles falling freely from rest at infinity in the background of a Sen black hole. Compared with the extremal KN black hole, an extremal Sen black hole can always act as a particle accelerator to an arbitrarily high energy with two conditions: (1) spin $a\neq 0$ and (2) one of the colliding particles has critical angular momentum $l_{\text{c}}=2$. For a nonextremal black hole, we find that an unlimited CM energy requires that one of the colliding particles has the critical angular momentum $l'_{\text{c}}=2 r_{+}/a$. However, the particles with $l=l'_{\text{c}}$ could not approach the black hole from outside of the horizon through free fall. Thus the CM energy is finite. Recently Grib and Pavlov
suggested that, an unlimited CM energy for a nonextremal black hole can be obtained when the
multiple scattering is considered \cite{Grib2,Grib4}. Motivated by this idea, we in this paper also consider the case that the colliding particles (may be produced from the multiple scattering near the horizon) start at some radiuses near the black hole and then fall into the black hole. However, the result shows that there always exists a forbidden band near the horizon for the particles with $l=l'_{\text{c}}$. So, these particles could not approach the horizon of the black hole from the outside. Thus, the CM energy is still limited for the non-extremal Sen black hole. We neglect the effects of gravitational radiation in the paper.

The paper is organized as follows. In section \ref{Review}, we first
briefly review the Sen black hole solution. In section \ref{Geodesics},
the geodesic and orbit equations for the particles in the background of
a Sen black hole are studied. With the effective potential method, we
discuss the range of the angular momentum, among which the particles can reach the horizon and fall
into the black hole. In section \ref{Center}, we study the CM energy
$E_{\text{cm}}$ for the collision taking place at the degenerate
horizon of an extremal black hole and at the outer
horizon of a nonextremal Sen black hole. The final section is
devoted to a brief summary.

\section{Review of the Sen black hole solution}
\label{Review}

In this section, we would like to give a brief introduction to the
Sen black hole solution, which is described by the four dimensional
effective action of the heterotic string theory:
\begin{eqnarray}
  S=-\int d^{4}x\sqrt{-\mathcal{G}}e^{-\Phi}\bigg(-\mathcal{R}
       +\frac{1}{12}\mathcal{H}^{2}
       -\mathcal{G}^{\mu\nu}\partial_{\mu}\Phi\partial_{\nu}\Phi
       +\frac{1}{8}\mathcal{F}^{2}\bigg),
       \label{action}
\end{eqnarray}
where $\Phi$ is the dilaton field and $\mathcal{R}$ is the scalar
curvature,
$\mathcal{F}^{2}=\mathcal{F}_{\mu\nu}\mathcal{F}^{\mu\nu}$ with the field strength
$\mathcal{F}_{\mu\nu}=\partial_{\mu}\mathcal{A}_{\nu}-\partial_{\nu}\mathcal{A}_{\mu}$
corresponds to the Maxwell field
$\mathcal{A}_{\mu}$, and
$\mathcal{H}^{2}=\mathcal{H}_{\mu\nu\rho}\mathcal{H}^{\mu\nu\rho}$
with $\mathcal{H}_{\mu\nu\rho}$ given by
\begin{eqnarray}
  \mathcal{H}_{\mu\nu\rho}&=&\partial_{\mu}\mathcal{B}_{\nu\rho}
                 +\partial_{\nu}\mathcal{B}_{\rho\mu}
                 +\partial_{\rho}\mathcal{B}_{\mu\nu}
                 -\frac{1}{4}\bigg(\mathcal{A}_{\mu}\mathcal{F}_{\nu\rho}
                 +\mathcal{A}_{\nu}\mathcal{F}_{\rho\mu}
                 +\mathcal{A}_{\rho}\mathcal{F}_{\mu\nu}\bigg),\label{Hmunurho}
\end{eqnarray}
where the last term in (\ref{Hmunurho}) is the gauge
Chern-Simons term. $\mathcal{G}_{\mu\nu}$ appeared in
(\ref{action}) are the covariant components of the metric in the
string frame, which are related to the Einstein metric by
$g_{\mu\nu}=e^{-\Phi}\mathcal{G}_{\mu\nu}$. The Einstein metric, the
non-vanishing components of $A_\mu$, $B_{\mu\nu}$ and the
dilaton field read \cite{Sen}:
\begin{eqnarray}
  ds^{2}&=&-\bigg(\frac{\Delta-a^{2}\sin^{2}\theta}{\Sigma}\bigg)dt^{2}
         +\frac{\Sigma}{\Delta}dr^{2}
         -\frac{4\mu ar\cosh^{2}\alpha\sin^{2}\theta}{\Sigma}dtd\phi
         +\Sigma d\theta^{2}
         +\frac{\Xi\sin^{2}\theta}{\Sigma}d\phi^{2},\label{metric}~~~~~~\\
  \mathcal{A}_{t}&=&\frac{\mu r\sinh 2\alpha}{\sqrt{2}\Sigma},
  \;\;\;\mathcal{A}_{\phi}=\frac{\mu a r\sinh 2\alpha\sin^{2}\theta}{\sqrt{2}\Sigma},\\
  \mathcal{B}_{t\phi}&=&\frac{2a^{2}\mu r\sin^{2}\theta\sinh^{2}\alpha}{\Sigma},
  \;\Phi=-\frac{1}{2}\ln \frac{\Sigma}{r^{2}+a^{2}\cos^{2}\theta},
\end{eqnarray}
where the metric functions are given by
\begin{eqnarray}
  \Delta&=&r^{2}-2\mu r+a^{2},\\
  \Sigma&=&r^{2}+a^{2}\cos^{2}\theta+2\mu r\sinh^{2}\alpha,\\
  \Xi&=&\bigg(r^{2}+2\mu r\sinh^{2}\alpha+a^{2}\bigg)^{2}
            - a^{2}\Delta\sin^{2}\theta.
\end{eqnarray}
The parameters $\mu$, $\alpha$ and $a$ are related to the physical
mass $M$, the charge $Q$ and the angular momentum $J$ by
\begin{eqnarray}
  M=\frac{\mu}{2}(1+\cosh 2\alpha),\;\;
  Q=\frac{\mu}{\sqrt{2}}\sinh^{2}2\alpha,\;\;
  J=\frac{a\mu}{2}(1+\cosh 2\alpha).\label{equation}
\end{eqnarray}
Solving eq. (\ref{equation}), we can obtain
\begin{eqnarray}
 \sinh^{2}\alpha=\frac{Q^{2}}{2M^{2}-Q^{2}},\;\;
 \mu=M-\frac{Q^{2}}{2M}.\label{relation}
\end{eqnarray}
Then the parameters $\alpha$ and $\mu$ in the metric (\ref{metric})
can be eliminated. For a nonextremal black hole, there are two
horizons. They are both determined by $\Delta(r)=0$ and are given by
\begin{eqnarray}
 r_{\pm}=M-\frac{Q^{2}}{2M}\pm\sqrt{\bigg(M-\frac{Q^{2}}{2M}\bigg)^{2}-a^{2}},
 \label{horizons}
\end{eqnarray}
where $r_{+}$ is the outer horizon and $r_{-}$ is the inner one.
Obviously, the extremal Sen black hole requires
\begin{eqnarray}
 Q^{2}=2M(M-a).
\end{eqnarray}
Setting $Q$ and $a$ to zero, respectively, we can obtain the maximum
values for them. Thus, we obtain the ranges for $a$ and $Q$, which
are
\begin{eqnarray}
 && 0\leq a\leq M,\\
 && 0\leq Q\leq \sqrt{2}M.
\end{eqnarray}
Here, both the parameters $a$ and $Q$ are thought to be positive.
For an extremal black hole, the two horizons coincide with each
other and the degenerate horizon locates at $r_{\text{ex}}=a$. The
area of the outer horizon with the metric given in (\ref{metric}) is
\begin{eqnarray}
 A=8\pi M
 \bigg(M-\frac{Q^{2}}{2M}+\sqrt{\bigg(M-\frac{Q^{2}}{2M}\bigg)^{2}-a^{2}}\bigg).
\end{eqnarray}
Thus the entropy of the Sen black hole is $S=\frac{A}{4}$. The
angular velocity $\Omega$, Hawking temperature $T$ and electric
potential $V_{+}$ at the outer horizon are
\begin{eqnarray}
 \Omega&=&\frac{J}{2M^{2}}
            \frac{1}{M-\frac{Q^{2}}{2M}+\sqrt{(M^{2}-\frac{Q^{2}}{2M})^{2}-a^{2}}},\\
 T&=&\frac{\sqrt{(2M^{2}-Q^{2})^{2}-4J^{2}}}
            {4\pi M(2M^{2}-Q^{2}+\sqrt{(2M^{2}-Q^{2})^{2}-4J^{2}})},\\
 V_{+}&=&\frac{Q}{2M}.
\end{eqnarray}
It is easy to check that all the quantities satisfy the first law of
black hole thermodynamics, $TdS=dM-\Omega dJ-V_{+}dQ$. With the help
of (\ref{relation}), we can get the inverse metric
\begin{eqnarray}
 (\partial_s)^{2}&=&
      -\frac{\Xi\Sigma}{\Delta\Xi+a^{2}\sin^{2}\theta(4M^{2}r^{2}-\Xi)}(\partial_t)^{2}
      -\frac{4aMr\Sigma}
      {\Delta\Xi+a^{2}\sin^{2}\theta(4M^{2}r^{2}-\Xi)}(\partial_t)(\partial_
      \phi)\nonumber\\
      &+&\frac{\Delta}{\Sigma}(\partial_r)^{2}
      +\frac{1}{\Sigma}(\partial_\theta)^{2}
      +\frac{\Sigma(\Delta\csc^{2}\theta-a^{2})}
          {\Delta\Xi+a^{2}\sin^{2}\theta(4M^{2}r^{2}-\Xi)}(\partial_\phi)^{2}.
\end{eqnarray}
And the metric functions can be rewritten as
\begin{eqnarray}
  \Delta&=&a^{2}+\frac{r}{M}\bigg(Q^{2}+Mr-2M^{2}\bigg),\\
  \Sigma&=&\frac{r}{M}\bigg(Q^{2}+Mr\bigg)+a^{2}\cos^{2}\theta,\\
  \Xi&=&\bigg(a^{2}+\frac{r}{M}(Q^{2}+Mr)\bigg)^{2}
          -a^{2}\Delta\sin^{2}\theta.
\end{eqnarray}
Here, we can explicitly see that the Sen black hole is characterized
by three parameters, mass $M$, charge $Q$ and spin $a$. The Sen
black hole solution will describe a Gibbon-Maeda (GM) black hole
with $a=0$ or describe a Kerr black hole with $Q=0$.

\section{Geodesics and orbit equation}
\label{Geodesics}

\subsection{First-order geodesic equations for particles}

The motion of a particle in the background of a Sen black hole is described by the geodesic equation
\begin{eqnarray}
  \frac{d^{2}x^{\mu}}{d\lambda^{2}}
    +\Gamma^{\mu}_{\;\;\nu\sigma}
    \frac{dx^{\nu}}{d\lambda}\frac{dx^{\sigma}}{d\lambda}=0, \label{GEQ}
\end{eqnarray}
where $\lambda$ and $\Gamma^{\mu}_{\;\;\nu\sigma}$ are the affine parameter and the Christoffel symbols of the background geometry. The parameter $\lambda$ relates to the proper time by $\tau=\delta \lambda$ ($\delta=-1$,0,1 for spacelike geodesics,
null geodesics and timelike geodesics, respectively). Allied
with initial conditions, the geodesic equation (\ref{GEQ}) has a unique
solution. However, it is difficult to solve the equation directly. Luckily, one can use the Lagrangian approach to the problem. In this subsection, we will not give the details for the derivation of the equations of motion for a particle and we refer the reader to the previous work \cite{Blaga,Hioki,Houri}. In \cite{Blaga}, the radial geodesics around a Kerr-Sen black hole were studied. The null geodesics and
photon capture in the Sen black hole were investigated in \cite{Hioki}. For a charged particle, the equations of motion were studied in \cite{Houri}. In this paper, we mainly deal with the uncharged massive particle ($\delta$=1) and the equations of motion are \cite{Blaga,Hioki,Houri}

\begin{eqnarray}
 \frac{dt}{d\tau}&=&\frac{1}{\Delta\Sigma}(\Xi E-2Marl),\label{tequation}\\
 \frac{dr}{d\tau}&=&\sigma_{r}\frac{\sqrt{\Re}}{\Sigma},\label{Rad}\\
 \frac{d\theta}{d\tau}&=&\sigma_{\theta}\frac{\sqrt{\Theta}}{\Sigma},
          \label{thetaequation}\\
 \frac{d\phi}{d\tau}
     &=&\frac{1}{\Delta\Sigma}\bigg(2MarE+l\csc^{2}\theta(\Sigma-2Mr)\bigg)\label{phiequation}
\end{eqnarray}
with ${\Re}$ and $\Theta$ given by
\begin{eqnarray}
 {\Re}&=&\bigg[\big(r(r+Q^{2}/M)+a^{2}\big)E-al\bigg]^{2}
            -\Delta(r(r+Q^{2}/M)+\mathcal{K}),\\
 \Theta&=&\mathcal{K}-(l-aE)^{2}
            -\bigg[a^{2}(1-E^{2})+l^{2}\csc^{2}\theta\bigg]\cos^{2}\theta,
\end{eqnarray}
where the constants $E$ and $l$ are the conservation of energy and
orbital angular momentum per unit mass of the motion and they
correspond to the Killing fields $\partial_{t}$ and
$\partial_{\phi}$, respectively. The variable $\mathcal{K}$ is a separation constant.
The sign functions $\sigma_{r}=\pm$ and $\sigma_{\theta}=\pm$ are
independent from each other. Eqs.
(\ref{tequation})-(\ref{phiequation}) are the first-order geodesic
equations for a particle. With these equations, we could obtain the 4-velocity of a particle in the black hole background, which will be
used in our calculation for the CM energy of the collision in the next
section.

\subsection{Radial motion and effective potential on the equatorial plane}
\label{Radialpotential}

Here, we would like to study the radial motion of the particle
falling freely from rest at infinite in the background of a Sen
black hole. From the last subsection, we can see that the geodesic
line of a particle in the Sen metric is completely determined
by the first-order geodesic equations
(\ref{tequation})-(\ref{phiequation}). In this subsection, we
would like to study the radial motion of the particle on the
equatorial plane and then determine the range of the angular
momentum, among which the particle can fall into the black hole
\footnote{Here, ``the particle falls into the black hole" means that
the particle can fall into the outer horizon of the black hole.}.
For simplicity, we take $E=1$. The mass of the
black hole is also set to $M=1$.

We should keep in mind that the particle falls into the black hole
with a spiral orbit like that a small boat spirals into a big whirlpool.
And there exist two branches for the particle to fall into the black
hole. One branch of trajectory is for the particle with large
orbital angular momentum. The particle will spiral into a circular
orbit at some radiuses (larger than the radius of horizon), taking a
divergent proper time to do so. Another branch corresponds to the
particle with small orbital angular momentum, which can begin at some
radius $r$ and then spirals into the black hole subsequently. Thus,
there must exist a range of the orbital angular momentum, among which
the particle can reach the horizon of the black hole with no turning
point. On the other hand, since the black hole is a rotating one,
the absolute values of the maximum and
minimum critical angular momenta are not equal.

In fact, the critical angular momentum and the corresponding radius
can be found from the effective potential. Then the
range of the angular momentum for the particles can be determined from it. In this paper, we mainly focus on the
equatorial plane of the black hole, so we take the choice
$\theta=\frac{\pi}{2}$ throughout this paper. Then we rewrite
(\ref{Rad}) as
\begin{eqnarray}
 \frac{\dot{r}^{2}}{2}+V_{\text{eff}}=0,
\end{eqnarray}
where $V_{\text{eff}}$ is the effective potential and it is given by
\begin{eqnarray}
 V_{\text{eff}}&=&-\frac{\Re}{2\Sigma^{2}}\nonumber\\
        &=&-\frac{2r^{2}+(2Q^{2}-l^{2})r+(2a^{2}-4al-l^{2}(Q^{2}-2))}
                 {2r(Q^{2}+r)^{2}}.
\end{eqnarray}
The limiting values of the effective potential $V_{\text{eff}}$ at
infinity and at the horizon are
\begin{eqnarray}
 V_{\text{eff}}(r=\infty)&=&0,\\
 V_{\text{eff}}(r=r_{+})&=&\frac{2 \bigg(8 a l+b l^2-4 b+(l^2+4)(Q^2-2)\bigg)}
        {(b+2-Q^2)(b+2+Q^2)^2},
\end{eqnarray}
with $b=\sqrt{(2-Q^2)^2-4 a^2}$. As expected, the effective
potential $V_{\text{eff}}$ vanishes at infinity despite the values
of $a$ and $Q$. The values of the critical radius and the angular
momentum are determined by the conditions
\begin{eqnarray}
 V_{\text{eff}}=0 \quad \mbox{and} \quad \partial_{r}V_{\text{eff}}=0.
\end{eqnarray}
Solving these equations, we get four solutions:
\begin{eqnarray}
 L_{1,2}&=&-2\pm \sqrt{4-2Q^{2}+4a},\;\;\;r_{1,2}=2-Q^{2}+a\mp\sqrt{4-2Q^{2}+4a};\\
 L_{3,4}&=&2\mp \sqrt{4-2Q^{2}-4a},\;\;\;\;\;\;
 r_{3,4}=2-Q^{2}-a\mp\sqrt{4-2Q^{2}-4a}.
\end{eqnarray}
With the numerical calculation, we have $r_{1}\leq r_{3}<
r_{+}\leq r_{4}\leq r_{2}$ for arbitrary values of spin $a$ and
charge $Q$. Note that $0<r_{3}<r_+$. So, a particle with $l=L_{3}$
can fall into the black hole. The two positive radiuses $r_{2}$ and
$r_{4}$ are larger than the horizon $r_{+}$. Thus the angular
momenta correspond to the two positive radiuses $r_{2}$ and $r_{4}$
can determine the range ($L_2, ~L_4$), among which the particle can
approach the horizon and then fall into the black hole. The two
critical angular momenta have different sign, i.e., $L_{2}<0$ and
$L_{4}>0$. It is also found that they satisfy the relation
$|L_{2}|\geq|L_{4}|$ for arbitrary spin $a$ and charge $Q$. When
$a=Q=0$, one gets $|L_{2}|=|L_{4}|=4$, which is just the case for
the Schwarzschild black hole. It is also clear that, in the case of
$Q=0$, the result for the Kerr black hole \cite{Jacobson} will be
recovered.

\begin{figure*}
\centerline{\subfigure[]{\label{Rnonextremal}
\includegraphics[width=8cm,height=6cm]{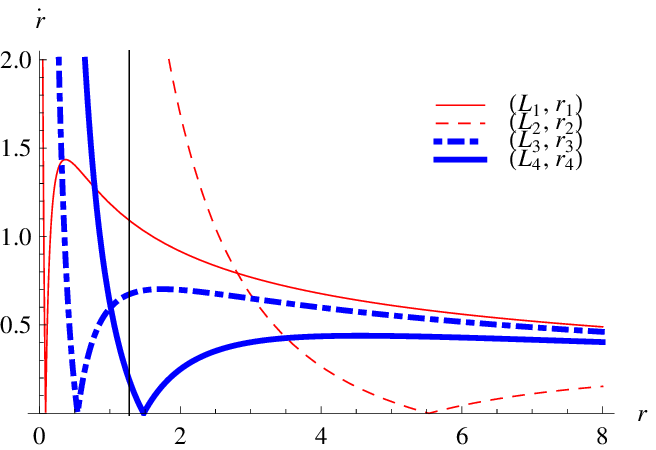}}
\subfigure[]{\label{Rextremal}
\includegraphics[width=8cm,height=6cm]{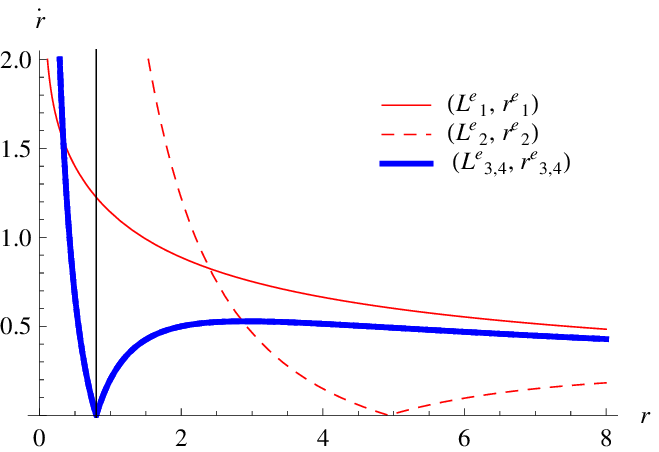}}}
\caption{The variation of $\dot{r}$ vs radius $r$ for different
values of critical angular momentum and the vertical lines denote
the locations of the outer horizon. (a) for a nonextremal Sen black
hole with $(a=0.9,\;Q=0.3)$ and (b) for an extremal Sen black hole with
$(a=0.9,\;Q=\sqrt{2-2a})$.} \label{PRad}
\end{figure*}

For an extremal Sen black hole, i.e., $Q^{2}=2(1-a)$, we have the
following solutions for the critical angular momenta and radiuses:
\begin{eqnarray}
 &&L_{1,2}^{\text{e}}=-2\pm 2\sqrt{2a},\;\;\;r_{1,2}^{\text{e}}=3a\mp 2\sqrt{2a};\\
 &&L_{3,4}^{\text{e}}=2,\;\quad\quad\quad\quad\quad r_{3,4}^{\text{e}}=a.
\end{eqnarray}
Note that $(L_{3},\;r_{3})$ is consistent with $(L_{4},\;r_{4})$ for
an extremal black hole. Because the horizon of an extremal black
hole locates at $r=r_{\text{ex}}=a$, the critical radiuses
$r_{3,4}^{\text{e}}=a$ coincide with the horizon $r_{\text{ex}}$. It can be
seen that $3a-2\sqrt{2a}\leq a$ (or $r_{1}^{\text{e}}\leq r_{\text{ex}}$),
so a particle with angular momentum $l=L_{1}^{\text{e}}=-2+ 2\sqrt{2a}$ can
fall into the horizon of the black hole. Thus we obtain the range of
the angular momentum:
\begin{eqnarray}
 -2-2\sqrt{2a}\leq l\leq 2 \quad \mbox{or} \quad
 L_{2}^{\text{e}}\leq l\leq L_{3,4}^{\text{e}}, \label{range}
\end{eqnarray}
among which the particle can fall into an extremal black hole.

In order to understand the explicit behavior of the radial equation
$\frac{dr}{d\tau}$ with respect to $r$, we plot it for the particles
with different critical angular momenta for the extremal and
nonextremal Sen black holes with the fixed spin $a=0.9$ in figure
\ref{PRad}. The case for the nonextremal black hole is depicted in
figure \ref{Rnonextremal}. We can see that the four critical radiuses
satisfy the relation $r_{1}\leq r_{3}<r_{+}\leq r_{4}\leq r_{2}$ and
all of the critical radiuses are well separated from the horizon
$r_{\text{ex}}=a=0.9$. While for the extremal black hole showed in
figure \ref{Rextremal}, we observe that the critical radiuses
$r^{\text{e}}_{3}$ and $r^{\text{e}}_{4}$ are equal and both coincide with the
horizon $r_{\text{ex}}$.

\section{Center-of-mass energy for collision in the background of a Sen black hole}
\label{Center}

In the above section, we have analyzed the range of the angular momentum,
among which the particle can reach the horizon. In other words, if the
angular momenta of the colliding particles are in the range, then the
collision can take place at the horizon of the black hole. So, based
on the results, we in this section will study the collision of two uncharged
particles (e.g. the massive cold dark matter particles) in the background of a Sen black hole. Here we consider
that two uncharged particles with the same rest mass $m_0$ are at
rest at infinity ($E=m_0$) and then they approach the black hole and
collide at radius $r$. For the sake of simplicity, we assume that
the two particles have angular momenta $l_{1}$ and $l_{2}$,
respectively. The energy in the center-of-mass frame for the
collision in the background of a Sen black hole is \cite{Banados}
\begin{eqnarray}
 E_{\text{cm}}=\sqrt{2}m_{0}\sqrt{1-g_{\mu\nu}u_{(1)}^{\mu}u_{(2)}^{\nu}},
  \label{energy}
\end{eqnarray}
where $u_{(1)}^{\mu}$ and $u_{(2)}^{\nu}$ are the 4-velocities of
the two particles. With the help of the first-order geodesic
equations (\ref{tequation})-(\ref{phiequation}), we obtain the
4-velocity for a particle on the equatorial plane
\begin{eqnarray}
 u^{t}&=&\frac{a^2 (Q^2+r+2)+r(Q^2+r)^2-2 a l}
                {(Q^2+r) [a^2+r(Q^2+r-2)]},\\
 u^{r}&=&\frac{\sqrt{r[2a^2-4al-l^2 (Q^2+r-2)+2r(Q^2+r)]}}{r (Q^2+r)},\\
 u^{\theta}&=&0,\\
 u^{\phi}&=&\frac{2a+l(Q^2+r-2)}{(Q^2+r)[a^2+r(Q^2+r-2)]},
\end{eqnarray}
where we have taken $E=1$. Note that the 4-velocity closely depends
on the spin $a$ and charge $Q$ of the black hole, also on the
angular momentum $l$ per unit mass of the particle.

\subsection{Extremal black hole}
\label{Extremalblackhole}

Note that for an extremal black hole, the horizon always locates at
$r=r_{\text{ex}}=a$ and the spin $a$ and charge $Q$ satisfy the
relation $Q^{2}=2(1-a)$. Then, we get the
4-velocity for a particle in the background of an extremal black
hole:
\begin{eqnarray}
 u^{\mu}=&&\bigg(\frac{r(r+2)^2+a^2 (5 r+4)-2a^3-2a(L+2r(r+2))}
          {(a-r)^2 (r-2a+2)},\nonumber\\
     &&-\frac{\sqrt{r(2a(a+(l-2)l)-r (4a+L^2-4)+2 r^2)}}
       {r (r-2a+2)},
        0,
     \frac{l r-2a(l-1)}{(a-r)^2 (r-2a+2)}\bigg).
\end{eqnarray}
Using (\ref{energy}), we get the CM energy for two colliding
particles at radius $r$ in the background of an extremal Sen black hole
\begin{eqnarray}
 \bigg(\frac{E_{\text{cm}}}{\sqrt{2}m_{0}}\bigg)^{2}=\frac{K}{(r-2a+2) (r-a)^2}, \label{Nonen}
\end{eqnarray}
with $K$ is given by
\begin{eqnarray}
 K=&&4 r+2 (r-2a+3)(a-r)^2-l_1 l_2 (r-2 a)
    -2 a (l_1+l_2)\nonumber\\
    &&-\sqrt{l_1^2 (2 a-r)+4(r-al_1)
       +2(r-a)^2} \sqrt{l_2^2 (2a-r)+4(r-a l_2)+2 (r-a)^2}.~~~
\end{eqnarray}
Note that $(r-2a+2)$ is always positive for an arbitrary spin $a$.
So, we only observe a double
zero at the horizon $r_{\text{ex}}=a$ for the denominator in (\ref{Nonen}). However, the numerator also has a double zero at that point. So the CM energy may
be limited for generic values of $l_{1}$ and $l_{2}$ at the horizon. Here, we
give the limiting values of the CM energy at
the horizon:
\begin{eqnarray}
 E_{\text{cm}}(r=r_{\text{ex}})=2m_{0}\sqrt{1+\frac{(l_{1}-l_{2})^{2}}
             {2a(l_{1}-l_{\text{c}})(l_{2}-l_{\text{c}})}}, \label{ECME}
\end{eqnarray}
with the critical angular momentum
\begin{eqnarray}
 l_{\text{c}}=2. \label{critical2}
\end{eqnarray}
For $l_{1}=l_{2}$, we get $E_{\text{cm}}=2m_{0}$ as expected.
Obviously, the CM energy $E_{\text{cm}}$ is finite for generic
values of $l_{1}$ and $l_{2}$. However, when $l_{1}=l_{\text{c}}$ or $l_{2}=l_{\text{c}}$, the CM energy
$E_{\text{cm}}$ will be unlimited. Thus, an extremal Sen black hole
can act as a particle accelerator to an arbitrarily high energy.
However, we need to make sure that the collision can take place at the horizon. In
fact, the result is obvious since $l_{\text{c}}$ is in the range
(\ref{range}), among which the particle can reach the horizon of an
extremal black hole. So, if the angular momentum of another
colliding particle is in the range (\ref{range}), the collision can
take place and an arbitrarily high CM energy will be approached.

It is known that, for an extremal KN black hole \cite{Wei}, in order to
obtain an unlimited CM energy, one of the colliding particles should have the critical angular momentum $l_{\text{c}}=\frac{1+a^{2}}{a}$. It can be seen that
when the spin $a$ of the black hole approaches 0, $\frac{1+a^{2}}{a}$ is
divergent, which will lead to a restriction that only the fast
rotating extremal black holes can act as a particle accelerator to an
arbitrarily high energy. However, for an extremal Sen black hole,
we can see that the critical angular momentum $l_{\text{c}}$
(\ref{critical2}) is a constant and independent of the parameters of
the black hole. So there is no restriction on the spin $a$ of the
extremal Sen black hole, which means that an arbitrary extremal
black hole can serve as a particle accelerator to an arbitrarily high
energy.

\begin{figure*}
\centerline{\subfigure[]{\label{Eextremal}
\includegraphics[width=8cm,height=6cm]{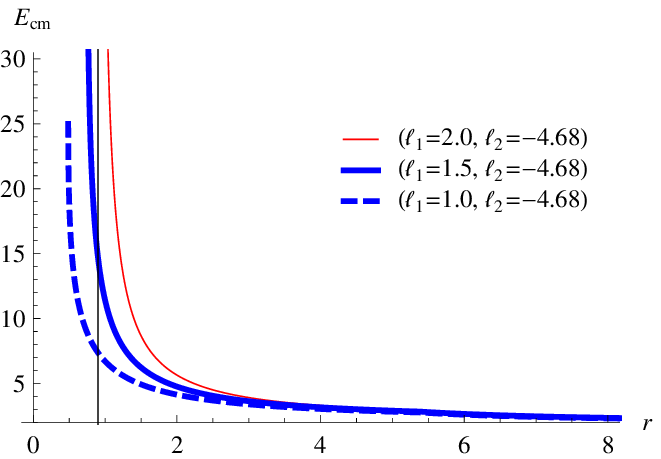}}
\subfigure[]{\label{LEextremal}
\includegraphics[width=8cm,height=6cm]{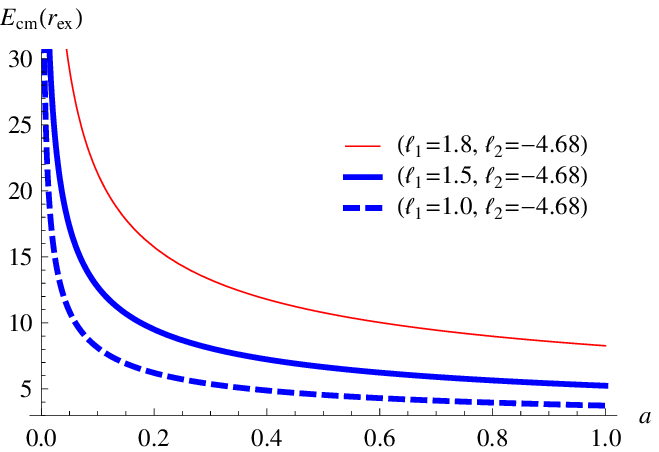}}}
\caption{The behavior of CM energy for an extremal Sen black hole
with $m_{0}=1$. (a) the CM energy $E_{\text{cm}}$ vs radius $r$
with fixed spin $a=0.9$ for different combinations of $l_{1}$ and
$l_{2}$. The vertical line denotes the location of the outer
horizon. (b) the CM energy $E_{\text{cm}}$ at horizon vs spin $a$
for different combinations of $l_{1}$ and $l_{2}$. }
\label{PLEextre}
\end{figure*}

Note that when the spin $a$ of the extremal Sen black hole
approaches its maximal value 1, we can express the CM energy
(\ref{ECME}) in the form
\begin{eqnarray}
 E_{\text{cm}}(r=r_{\text{ex}})=\sqrt{2}m_{0}
       \sqrt{\frac{l_{2}-2}{l_{1}-2}+\frac{l_{1}-2}{l_{2}-2}},
\end{eqnarray}
which is just the result for the extremal Kerr black hole given in
\cite{Banados}. From (\ref{ECME}), it is also clear that, when the spin $a$ approaches 0, we will obtain an unlimited CM energy $E_{\text{cm}}$ without
additional restriction on the angular momentum of the colliding particles except $-2\leq l\leq 2$. Since
the extremal Sen black hole with spin $a=0$ will be an extremal
charged GM black hole, this result implies that an extremal
charged GM black hole may also play the role of a particle
accelerator to an arbitrarily high energy. So the charged GM black hole
will share the same property as the rotating black hole to
accelerate particles. However, it is not the case. It is easy to
check that, for the extremal charged GM black hole, the radius of the
horizon will reduce to zero. Then there will be a naked singularity
at $r=0$ with no horizon around it. This case is forbidden according to the Penrose cosmic censorship conjecture that the singularity should be surrounded by a
horizon.

We plot in figure \ref{Eextremal} $E_{\text{cm}}$ vs $r$ for various
values of $l_{1}$ and $l_{2}$ with the fixed $a=0.9$. From the figure,
we can see that $E_{\text{cm}}$ blows up at the horizon when
$l_{1}=2$, while for other values of $l_{1}$, the $E_{\text{cm}}$
is finite. We can also find that when $r>4.0$,
the CM energy $E_{\text{cm}}$ is almost equal for the different
combinations of $l_{1}$ and $l_{2}$. The limiting values of
$E_{\text{cm}}$ at horizon for different spin $a$ of the black hole
are also plotted in figure \ref{LEextremal}. Note that the
$E_{\text{cm}}(r_{\text{ex}})$ will be unlimited when spin $a$
approaches zero. This case corresponds to the extremal charged GM
black hole and it is forbidden according to the Penrose cosmic
censorship conjecture from our above analysis.

One word to summarize this subsection is that an
extremal Sen black hole with $a\neq 0$ can serve as a particle
accelerator to an arbitrarily high energy if one of the colliding
particles has the critical angular momentum $l=2$.

\subsection{Non-extremal black hole}

As noted above, we show that an arbitrarily high CM energy can be
obtained when the collision takes place at the horizon of an
extremal Sen black hole with $l=l_{\text{c}}$ and $a\neq 0$.
However, this scenario is an idealized one for there are no
extremal astrophysical black holes. So our main
goal in this section is to study the collision of two particles in the background of
a nonextremal Sen black hole. Applying (\ref{energy}) to the case,
we get the CM energy
\begin{eqnarray}
 \bigg(\frac{E_{\text{cm}}}{\sqrt{2}m_{0}}\bigg)^{2}=\frac{H}{(r+Q^2) (r^{2}-r
   (2-Q^2)+a^2)}\label{nonextremalenergy}
\end{eqnarray}
with
\begin{eqnarray}
 H=&&2 r^3+2a^2(r+Q^2+1)-2r^2(1-2Q^2)\nonumber\\
   &&-2r Q^2(1-Q^2)-2a(l_1+l_2)
     -(r+Q^2-2)l_1 l_2 \nonumber\\
   &&-\sqrt{2(a-l_1)^2+(2 r-l_1^2)(r+Q^2)}
     \sqrt{2(a-l_2)^2+(2r-l_2^2) (r+Q^2)}.
\end{eqnarray}
Taking the limit of CM energy (\ref{nonextremalenergy}) at
$r=\infty$, we get $E_{\text{cm}}(r=\infty)=2m_{0}$, which is the
same as that the collision takes place in a flat spacetime. So does
the CM energy $E_{\text{cm}}$ in (\ref{Nonen}).

\begin{figure*}
\centerline{\subfigure[]{\label{Enextremalouter}
\includegraphics[width=8cm,height=6cm]{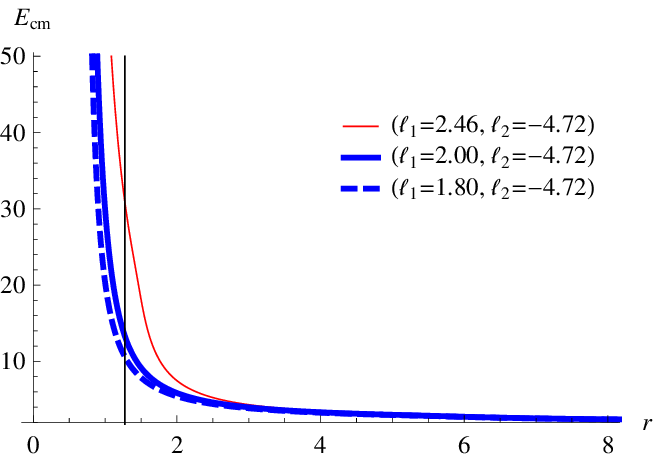}}
\subfigure[]{\label{LEnextremalouter}
\includegraphics[width=8cm,height=6cm]{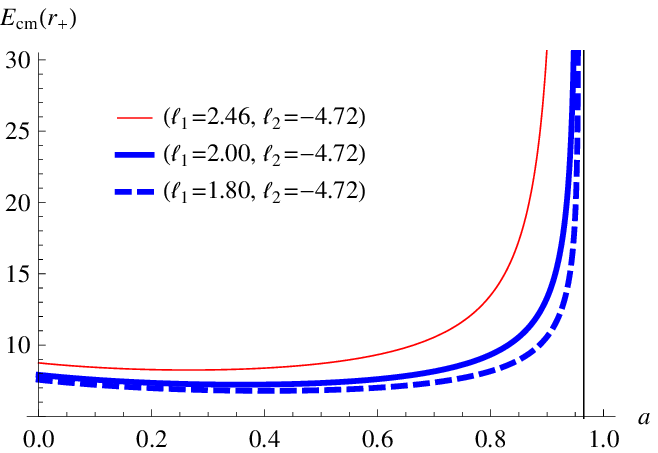}}}
\caption{(a) The CM energy $E_{\text{cm}}$ vs radius $r$ for
nonextremal black hole with $a$=0.9, $Q$=0.3. The vertical line
denotes the location of the outer horizon. (b) The CM energy
$E_{\text{cm}}$ at the outer horizon for different spin $a$ with
fixed charge $Q$=0.3. The vertical line denotes the location of the
maximum spin $a_{\text{max}}$ corresponding to the extremal black
hole.} \label{PEnextremalouter}
\end{figure*}

Next, we would like to consider the collision at the outer horizon
$r=r_{+}$. As shown in (\ref{horizons}), the outer horizon locates
at
$r_{+}=1-\frac{Q^{2}}{2}+\sqrt{\big(1-\frac{Q^{2}}{2}\big)^{2}-a^{2}}$.
After some calculations, we derive the limiting value of the CM
energy $E_{\text{cm}}$ at the outer horizon and which is given by
\begin{eqnarray}
 \frac{E_{\text{cm}}(r=r_{+})}{2m_{0}}= \sqrt{1+\frac{(l_1-l_2)^2}
 {2r_-(l_1-l'_{\text{c}})(l_2-l'_{\text{c}})}}\label{EEEEMMM2}
\end{eqnarray}
with the critical angular momentum
\begin{eqnarray}
 l'_{\text{c}}={2r_+}/{a}.
\end{eqnarray}
As expected, $E_{\text{cm}}(r=r_{+})=2m_{0}$ when the angular
momenta $l_{1}=l_{2}$. It is also clear that when $l_{1}=l'_{\text{c}}$ or
$l_{2}=l'_{\text{c}}$, we get an unlimited CM energy. It seems that we can
get an arbitrarily high CM energy if one of the colliding particles
has $l=l'_{\text{c}}$. However, we should make sure that particles with the
critical angular momentum $l'_{\text{c}}$ can reach the horizon. In section
\ref{Radialpotential}, we get the range ($L_{2},\;L_{4}$) for the angular momentum. If the angular momentum of a particle lies in the range, then the particle can reach the horizon and fall into the black hole. By numerical calculation, we
get $l'_{\text{c}}>L_{4}$ for arbitrary spin $a$ and charge $Q$, which
means that particles with $l'_{\text{c}}$ could not fall
into the black hole. Thus, for a nonextremal Sen black hole, the CM
energy is limited. The behavior of the CM energy $E_{\text{cm}}$ is
plotted in figure \ref{Enextremalouter} for a nonextremal black hole
with spin $a$=0.9 and charge $Q$=0.3. We can see that the CM energy
$E_{\text{cm}}$ for different combinations of $l_{1}$ and $l_{2}$ is
finite at the horizon. Figure \ref{LEnextremalouter} shows the CM energy
$E_{\text{cm}}$ at the outer horizon vs spin $a$ with fixed charge
$Q$=0.3. It is clear that the CM energy blows up at $a=0.9550$,
which is nothing but the spin of an extremal black hole with charge
$Q=0.3$. Therefore, it implies that the CM energy $E_{\text{cm}}$ is
unlimited for the extremal black hole, which exactly agrees with our
result obtained in subsection \ref{Extremalblackhole}.

The maximal CM energy for a nonextremal black hole can be obtained with
$l_{1}=L_{2}$ and $l_{2}=L_{4}$. It has a complicated form and we
will not list it here. However it is clear that the CM energy
depends on the charge $Q$ and spin $a$. Note that when a
nonextremal black hole approaches to the extremal one, the CM
energy tends to infinite. So, in order to obtain a high CM energy,
we will consider the case of a near-extremal black hole. First, we
define a small parameter $\epsilon=a_{\text{max}}-a$ with
$a_{\text{max}}=1-\frac{Q^{2}}{2}$. For a near-extremal
black hole, we have $\epsilon\ll 1$. With $l_{1}=L_{2}$ and $l_{2}=L_{4}$,
the maximal CM energy $E^{\text{max}}_{\text{cm}}$ can be approximated as
\begin{eqnarray}
 \frac{E^{\text{max}}_{\text{cm}}}{m_{0}}\sim
                 11.66\epsilon^{-1/2}+\mathcal{O}(\epsilon^{1/2}).
 \label{ECMlimitation}
\end{eqnarray}
Comparing with the result of Kerr black hole \cite{Jacobson}, i.e., $\frac{E^{\text{max}}_{\text{cm}}}{m_{0}}\sim 4.06\epsilon^{-1/4}$, we find that the CM energy for a Sen black hole grows faster than that for a Kerr black hole when $\epsilon\rightarrow 0$. The maximal CM energy (\ref{ECMlimitation}) will be used to discuss the back-reaction effects latter and we will see that the different $E^{\text{max}}_{\text{cm}}$ for the Kerr and Sen black holes will give different describes on the back-reaction effects. If the rest mass of the colliding particles is of about 1 GeV, then in order to obtain the Planck-scale energy
$E_{\text{Pl}}\sim 10^{19}$ GeV, we need
\begin{eqnarray}
 \epsilon\sim 10^{-36}.
\end{eqnarray}
The parameter $\epsilon$ is too small. So, it is very hard for the near-extremal black hole to be a
particle accelerator to Planck-scale energy.

\begin{figure*}
\centerline{
\includegraphics[width=10cm,height=6cm]{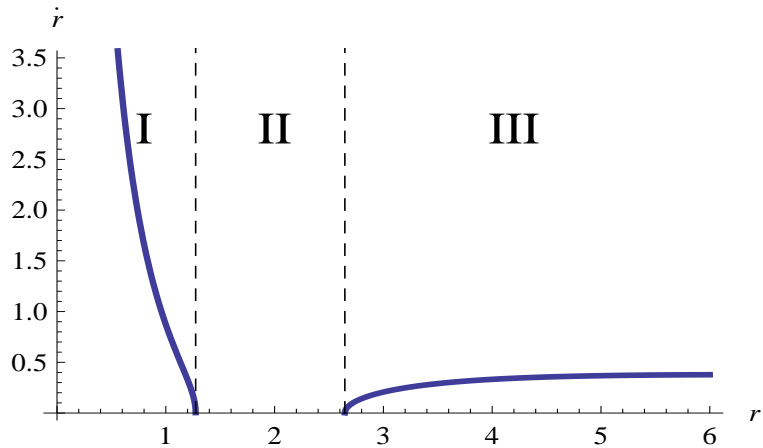}}
\caption{The variation of $\dot{r}$ vs radius $r$ for a particle
with critical angular momentum $l'_{\text{c}}$ in the background of a
nonextremal black hole with $a$=0.9 and $Q$=0.3.}
\label{PTcritical}
\end{figure*}

Here, we obtain a limited CM energy for the collision taking place
at the outer horizon of a nonextremal Sen black hole. However, we
should keep in mind that the colliding particles fall freely from
rest at infinity. If the colliding particles start at some radiuses near the black hole (we can think the particles are produced from the multiple scattering near the black hole as suggested by Grib and Pavlov \cite{Grib2,Grib4}), then what will happen? Is the CM energy divergent? Here, we will explore this problem in detail. As we mentioned above, in order to obtain an unlimited CM energy, one of
the colliding particles must have angular momentum
$l=l'_{\text{c}}$. So, in order to obtain an unlimited CM energy, the major
problem is that whether the particle with the critical angular momentum can reach the horizon. To find the answer, let us first
analyze the figure \ref{PTcritical}, which shows that the radial speed vs
$r$ for a particle with $l=l'_{\text{c}}$ in the background
of a nonextremal black hole with $a=0.9$ and $Q=0.3$ (for other
values of $a$ and $Q$, the shape does not change). From it, we can
see that the space is divided into three regions: region I, region
II and region III. The particle with $l=l'_{\text{c}}$ can
only exist in the region I or region III. And region II is a forbidden band
for the particle (where the effective potential $V_{\text{eff}}>0$). For the existence of region II, the particle could not
fall freely into the black hole from rest at infinity $(E=m_{0})$
\footnote{For the extremal case, there exists no region II, so the particle with the critical angular momentum can reach the horizon.}.
In region III, the particle can fall freely from rest at infinity
and end with the bound of region II and region III. The particle in
region I can be thought to be produced near the black hole from the
collision in the accretion disc and it has the same
parameters as the one falling freely from rest at infinity. If
the particle starts in region I, then it may provide a way
for the particles with critical angular momentum to reach the horizon
and then the collision can take place at the horizon. Thus, an
arbitrarily high CM energy can be obtained. However, with $l=l'_{\text{c}}$,
we always have the bounds:
\begin{eqnarray}
 r_{\text{I,II}}&=&r_{+},\label{RI}\\
 r_{\text{II,III}}&=&\frac{(4-a^{2}-2Q^{2})r_{+}-a^{2}(2+Q^{2})}{a^{2}},
\end{eqnarray}
where $r_{\text{I,II}}$ is the bound of the region I and II, $r_{\text{II,III}}$ is
the bound of the region II and III. We always have $r_{\text{I,II}}<r_{\text{II,III}}$ for a nonextremal black hole.
Eq.~(\ref{RI}) means that the bound between region I and II just lies the
location of the horizon. So, region I lies inside the black hole.
Thus, a particle with $l'_{\text{c}}$ could not approach the black hole through free fall. So the collision with an arbitrarily high
CM energy could not take place.

However, a particle with $l=l'_{\text{c}}-\delta$ can approach the
black hole from outside of the horizon. Then, with $l_{1}=l'_{\text{c}}-\delta$ and
$l_{2}=0$, one can obtain the approximate CM energy:
\begin{eqnarray}
 \frac{E^{\text{max}}_{\text{cm}}}{m_{0}}\sim
                 2.99\;\delta^{-1/2}+\mathcal{O}(\delta^{1/2}).
\end{eqnarray}
From it, we can see that, in order to obtain a high CM energy, the $\delta l$
must be very small. So, we come to the conclusion: the CM energy for a nonextremal black hole is limited no matter whether the colliding particles fall into the black hole from infinity or from the vicinity of the black hole.



In the last of this paper, we give a brief discussion on the back-reaction effects. It was shown in \cite{Berti} that, for the case of a Kerr black hole with mass $M_{\text{Kerr}}\sim 100\text{M}_{\odot}$, the collision of a single electron pair is enough to basically destroy the planckian accelerator. Here, we will generalize the discussion to the Sen black hole geometry. Neglecting gravitational radiation, when an extremal Sen black hole absorbs a pair of colliding particles with mass $m_{0}$, the dimensionless spin is reduced by $\epsilon=(a_{\text{max}}-a)\sim \frac{m_{0}}{M}(1+\frac{Q^{2}}{2M^{2}})$. Then after the first collision, an extremal black hole will become a nonextremal one and the new maximum allowed CM energy would be
\begin{eqnarray}
 E_{\text{cm}}\lesssim 10^{28}\cdot(m_{0}/1\text{MeV})^{1/2}(M/100\text{M}_{\odot})^{1/2}\; \text{GeV}.
\end{eqnarray}
For $m_{0}\sim 1$ MeV and $M\sim 100\text{M}_{\odot}$, we get $E_{\text{cm}}\sim 10^{28}$ GeV, which is clear above the Planck scale. This result is very different from that of the Kerr black hole, $E^{\text{Kerr}}_{\text{cm}}\lesssim 10^{12}\cdot(m_{0}/1\text{MeV})^{1/2}(M/100\text{M}_{\odot})^{1/2}$ GeV \cite{Berti},  where a single electron pair is enough to destroy the planckian accelerator. So this back-reaction effects has a weak effect on the CM energy for the Sen black hole geometry. The crucial difference between the two black holes is that the two CM energies have different asymptotic behaviors when the small parameter $\epsilon\rightarrow 0$ (i.e., $E^{\text{max}}_{\text{cm}}/m_{0}\sim 11.66\epsilon^{-1/2}$ for the Sen black hole and $E^{\text{max}}_{\text{cm}}/m_{0}\sim 4.06\epsilon^{-1/4}$ for the Kerr black hole). However, we have neglected the gravitational radiation in our analysis, which will significantly affect the geodesics when $\delta=1-l/l_{\text{c}} \ll 1$, because the total radiated energy $E_{\text{tot}}\sim -\log \delta$ as discussed in \cite{Berti}. Thus as the gravitational radiation is considered, we argue that the maximum allowed CM energy will be greatly reduced to below the Planck-scale. The detailed analysis of gravitational radiation should be carried out in the future.

\section{Summary}
\label{Conclusion}

In this paper, we have investigated the property of the CM energy
$E_{\text{cm}}$ for two uncharged colliding particles in the background of a
Sen black hole. To get the CM energy, we first studied
the explicit expression of the first-order geodesic equations for a
particle, from which the 4-velocity of a particle
is obtained. We also analyzed the range of
the angular momentum, among which the particle can reach the horizon and fall into the black hole. If the angular momentum of a particle
lies outside the range, it will not fall into the black hole and
will turn back at some radiuses. For the case, the collision will
not take place at the radiuses smaller than the turning point.
Considering the range of the angular momentum, we investigated the property of the collision taking place at the horizon of a Sen black hole. The results show
that an extremal black holes with spin $a\neq 0$ could serve as
particle accelerators to an arbitrarily high energy with the fine-tuning $l_1=2$ or $l_2=2$. However, it seems a pity that, for a
nonextremal black hole, the CM energy is always finite.
The CM energy for a near-extremal black hole was also considered. The
result shows that, in order to obtain the Planck-scale energy, the small
parameter $\epsilon\sim 10^{-36}$, which implies that it is very hard for a near-extremal black hole to be a particle accelerator of Planck-scale energy. Furthermore, we also discussed the case that the particle approaches the horizon from a finite radius and collides with another particle. However, we found that the particle with $l=l'_{\text{c}}$ could not approach the black hole from outside of the horizon. So, the collision with an arbitrarily high CM energy will not take place. Thus, for a nonextremal Sen black hole, the CM energy is always limited no matter where the colliding particles fall from. We also show that the back-reaction effects has a weak effect on the CM energy. This result is very different from that of the Kerr black hole \cite{Berti}, where the back-reaction effects is strong enough to inhibit further Planck-scale collisions after the first collision. However the gravitational radiation will affect the CM energy dramatically for the total radiated energy $E_{\text{tot}}\sim -\log \delta$ and the CM energy may be below the Planck-scale when this effect is included.

\section*{Acknowledgements}

This work was supported by the Program for New Century Excellent
Talents in University, the Huo Ying-Dong Education Foundation of Chinese Ministry of Education (No. 121106), the National Natural Science Foundation of
China (No. 10705013 and No. 11075065), and the Fundamental Research Funds for the
Central Universities (No. lzujbky-2009-54 and No. lzujbky-2009-
163).


\begin{thebibliography}{99}

\bibitem{Banados}
  M. Banados, J. Silk and S. M. West,
   {\em Kerr Black Holes as Particle Accelerators to Arbitrarily High
                                          Energy},
     Phys. Rev. Lett. \textbf{103}, 111102 (2009),
     [arXiv:0909.0169[hep-ph]].

\bibitem{Berti}
  E. Berti, V. Cardoso, L. Gualtieri, F. Pretorius and U. Sperhake,
   {\em Comment on "Kerr Black Holes as Particle Accelerators to
                       Arbitrarily High Energy"},
     Phys. Rev. Lett. \textbf{103}, 239001 (2009),
      [arXiv:0911.2243[gr-qc]].

\bibitem{Jacobson}
   T. Jacobson and T. P. Sotiriou,
               {\em Spinning Black Holes as Particle Accelerators},
     Phys. Rev. Lett. \textbf{104}, 021101 (2010),
      [arXiv:0911.3363[gr-qc]].

\bibitem{Thorne}
   K. S. Thorne,
                 {\em Disk Accretion onto A Black Hole. 2. Evolution Of The Hole},
     Astrophys. J. \textbf{191}, 507 (1974).

\bibitem{Lake}
   K. Lake,
           {\em Particle Accelerators inside Spinning Black Holes},
    Phys. Rev. Lett. \textbf{104}, 211102 (2010),
      [arXiv:1001.5463[gr-qc]];
   K. Lake,
           {\em Erratum: Particle Accelerators Inside Spinning Black Holes [Phys. Rev. Lett. 104, 211102 (2010)]},
    Phys. Rev. Lett. \textbf{104}, 259903(E) (2010).


\bibitem{Grib1}
   A. A. Grib and Yu. V. Pavlov,
{\em On Particle Collisions in the Gravitational Field of Black
     Hole},
      [arXiv:1001.0756[gr-qc]].


\bibitem{Grib2}
    A. A. Grib and Yu. V. Pavlov,
  {\em On the collisions between particles in the vicinity of rotating black holes},
     JETP Lett. \textbf{92} 125 (2010),
      [arXiv:1004.0913[gr-qc]].

\bibitem{Grib3}
   A. A. Grib and Yu. V. Pavlov,
{\em On particle collisions near Kerr's black holes},
      [arXiv:1007.3222[gr-qc]].


\bibitem{Zaslavskii}
    O. B. Zaslavskii,
{\em Acceleration of particles as universal property of rotating
                black holes},
  Phys. Rev. \textbf{D 82}, 083004 (2010),
      [arXiv:1007.3678[gr-qc]].

\bibitem{Wei}
    S. W. Wei, Y. X. Liu, H. Guo and C. E Fu,
{\em Charged spinning black holes as particle accelerators},
    Phys. Rev. \textbf{D 82}, 103005 (2010),
    [arXiv:1006.1056[hep-th]].

\bibitem{Grib4}
   A. A. Grib and Yu. V. Pavlov,
{\em On particles collisions near rotating black holes},
      [arXiv:1010.2052[gr-qc]].



\bibitem{Sen}
   A. Sen,
    {\em Rotating charged black hole solution in heterotic string
                        theory},
    Phys. Rev. Lett. \textbf{69}, 1006 (1992),
      [arXiv:hep-th/9204046].


\bibitem{Blaga}
   P. A Blaga and C. Blaga,
   {\em Bounded radial geodesics around a Kerr-Sen black hole},
    Class. Quantum. Grav. \textbf{18}, 3893 (2001).


\bibitem{Hioki}
   K. Hioki and U. Miyamoto,
   {\em Hidden symmetries, null geodesics, and photon capture in the Sen black hole},
    Phys. Rev. \textbf{D 78}, 044007 (2008),
      [arXiv:0805.3146[gr-qc]].


\bibitem{Houri}
  T. Houri, D. Kubiznak, C. M. Warnick and Y. Yasui,
    {\em Generalized hidden symmetries and the Kerr-Sen black hole},
    JHEP \textbf{1007}, 055 (2010),
      [arXiv:1004.1032[hep-th]].


\end{thebibliography}
\end{document}